\documentstyle[osa,epsf,eqsecnum,epsfig,graphics,cite]{revtex}

\renewcommand{\theequation}{\thesection.\arabic{equation}}
\newcounter{hran} 
\def\noi{\noindent}
\def\bea{\begin{eqnarray}} 
\def\eea{\end{eqnarray}} 
\def\beq{\begin{equation}} 
\def\eeq{\end{equation}} 
\begin{document}

\title {How does transverse (hydrodynamic) flow affect
jet-broadening and jet-quenching ?}

\author{R.~Baier$^{\rm 1}$\footnote{E-mail address:
    baier@physik.uni-bielefeld.de},
 A.~H.~Mueller $^{\rm 2}$\footnote{E-mail address:
    arb@physics.columbia.edu }
~and 
D.~Schiff$^{\rm 3}$\footnote{E-mail
address: Dominique.Schiff@th.u-psud.fr}}

\address{
$^{\rm 1}$ Physics Department, University of Bielefeld, D-33501 Bielefeld,
Germany\\[0pt]
$^{\rm 2}$Physics Department, Columbia University, New York, NY 10027, USA\\
$^{\rm 3}$LPT, $B\hat{a}t.~210$, Universit\'e Paris-Sud, F-91405 Orsay, France
}

\date{\today}

\maketitle

\begin{abstract}

We give the modification of formulas for  $p_{\perp}$-broadening 
and energy loss which are necessary to calculate parton interactions in 
a medium with flow. 
Arguments are presented leading to the conclusion that for
 large $p_{\perp}$-spectra
 observed in heavy ion collisions at RHIC, the influence of 
transverse flow on the determination of the "quenching power" of the  
produced medium is small.
This leaves open the question of the interpretation of data in a 
consistent perturbative framework.

\end{abstract}

\section{Introduction}
\hspace*{\parindent}
Energy loss of a high transverse momentum parton travelling
 in the hot and dense medium created in ultrarelativistic collisions 
has been the subject in recent years of intense investigation, reviewed in
\cite{Baier:2000mf,Kovner:2003zj,Jacobs:2004qv,Jacobs:2005tq,d'Enterria:2005hh,Salgado:2005zp}.
 With the aim of being able in particular to assert the existence of 
the QGP phase of matter \cite{RHIC}.

It has become clear however, that an important issue is
 the influence of the medium evolution on the radiative energy loss, when
following the BDMPS \cite{Baier:1996kr,Baier:1996sk,Baier:1998kq} - Zakharov
 \cite{Zakharov:1997uu,Zakharov:2004vm} -
Wiedemann \cite{Wiedemann:2000tf} approach.
 The medium cannot be described as being static \cite{Gyulassy:1993hr}~:
 longitudinal and transverse flow have to be considered
\cite{Baier:1998yf,Salgado:2002cd,Wang:2002ri}.
 They may alter substantially the distribution of matter before freeze out.
 The crucial parameter for energy loss is the local transport coefficient
 $\widehat{q}(\tau )$, related to the squared average transverse momentum 
transfer -- from the medium to the hard parton -- per unit length.

For an ideal QGP, and in a comoving coordinate system,
 one may relate locally $\widehat{q}$ with the energy
 density of the medium $\varepsilon$~: $\widehat{q} \simeq c~ \varepsilon^{3/4}$ 
with $c \simeq 2$ \cite{Baier:2002tc,Baier:2006fr}.

Recently from various authors 
\cite{Eskola:2004cr,Arleo:2002kh,Arleo:2006xb,Turbide:2005fk}
came the determination of $c$ from data, 
 taking into account Bjorken longitudinal expansion 
\cite{Bjorken} but neglecting
 transverse flow. In \cite{Eskola:2004cr} $c$ 
is found to be much larger than 2
 ($c > 8 \cdots 19$, $c\simeq 10$),  leading  to speculations about
 a strongly coupled QGP
\cite{Shuryak:2004cy,Gyulassy:2004zy,Lee:2005gw,Heinz:2005zg,Blau:2005pk,Riordan:2006df} 
.
A recent endeavour \cite{Renk:2005rq,Renk:2005ta}
implementing a strong non-Bjorken expansion including radial flow
\cite{Armesto:2004vz},
with a small  value of primordial transverse velocity
$v^i = 0.1$
 has led to a ``moderately optimistic'' 
scenario reducing $c = 10$ to $c=2$ compatible with perturbative 
estimates.

In the present note, we try to reformulate the problem.
We show first how to determine $\hat{q}$ in the presence of transverse
flow by applying a proper Lorentz boost.
We assume a realistic (ideal) hydrodynamical description for
heavy-ion collisions, based on longitudinal Bjorken expansion
\cite{Bjorken} and a
radial flow with vanishing initial velocity.
 
 We are led  to the conclusion 
 that the transverse flow  is a small effect and does not provide the 
solution to the perturbative/non-perturbative dilemma.

\section{Moving medium versus medium at rest~: geometry}
\hspace*{\parindent} The problem is to find the relationship between
 $\widehat{q}$ for a moving medium and $\widehat{q}_0$ (medium at rest),
 locally in space and time. We focus on central collisions.\par

We consider -- in the transverse plane $(xy)$ of a collision-- 
a medium moving with velocity $\vec{v}$ along the $y$ axis towards a
 parton which enters into it
at the origin of our coordinate system  at time $\tau = 0$, at point $A$,
 leaving it at point $B$ with $B^{\mu} = t(1,  \sin \theta ,
 \cos \theta , 0 )$ where we denote by $t$ the time spent by the
 parton in the medium. The dimension of the medium in the velocity 
direction being $L$, one finds
\beq
\label{1e}
t = {L \over v + \cos \theta } \ .
\eeq

\noi The medium at rest has length $\displaystyle{{L \over \sqrt{1 - v^2}}} 
\equiv L \ {\rm ch}\ \zeta$ defining $\zeta$ such that $v = {\rm th}\ \zeta$.
 Entering at point $A_0$, the parton leaves the medium at rest,
 at point $B_0$ such that as a result of the Lorentz boost, we find
\beq
\label{2e}
B_0^{\mu} = t \left ( {\rm ch}\ \zeta \left ( 1 + \cos \theta\
 {\rm th}\ \zeta \right ) , \sin \theta ,  \ {\rm ch}\ \zeta \left (
  \cos \theta +\  {\rm th}\ \zeta \right ) , 0 \right )
\eeq

\noi with
\beq
\label{3e}
A_0B_0 = {L \ {\rm ch}\ \zeta \over \cos \theta_0}
 \equiv t\ {\rm ch} \ \zeta \left ( 1 + \cos \theta\ {\rm th}\ \zeta \right )
\eeq

\noi The angle $\theta_0$ corresponds to the angle $\theta$, but
evaluated in the rest system of the medium, 
 implying
\beq
\label{4e}
\cos \theta_0 = {{\rm th}\ \zeta + \cos \theta \over 1 +
 \cos \theta\ {\rm th}\ \zeta}
\eeq

\noi and
\beq
\label{5e}
\sin \theta_0 = {\sin \theta \over {\rm ch}\ \zeta \left
 ( 1 + \cos \theta \ {\rm th}\ \zeta \right )} \ .
\eeq

\section{Transverse momentum broadening}

\subsection{Local geometry}
\hspace*{\parindent}
Let us define in the rest frame the initial momentum of the parton~:

\beq
\label{6e}
p_0^{\mu} = p_0 (1, \sin \theta_0, \cos \theta_0, 0)\ .
\eeq

\noi After travelling on a small distance $\delta \tau$ -- we want to extract
 a {\bf local} information -- and experiencing $p_{\bot}$-broadening
 and energy loss, the parton has a momentum $\overline{p}_0^{\mu}$~:
\beq
\label{7e}
\overline{p}_0^{\mu} = \left ( p_0 - \delta \varepsilon_0\right ) \left 
( 1, \sin \left ( \theta_0 + \delta \theta_0\right ) \cos \delta \varphi_0,
 \cos \left ( \theta_0 + \delta \theta_0\right ), \sin \left ( \theta_0 +
 \delta \theta_0\right ) \sin \delta \varphi_0  \right )
\eeq

\noi where $\delta \varphi_0$ is the azimuthal angle in the $zx$ plane.
 In the moving frame, the corresponding momenta are~:
\bea
\label{8e}
&&p^{\mu} = p \left ( 1 , \sin \theta ,  \cos \theta , 0\right )\\
&&\overline{p}^{\mu} = (p - \delta \varepsilon ) \left ( 1, \sin \left ( \theta
 + \delta \theta\right ) \cos \delta \varphi, \cos \left ( \theta +
 \delta \theta\right  ) , \sin \left ( \theta + \delta \theta\right )
 \sin \delta \varphi \right ) \ .
\label{9e}
\eea

\noi
The momenta in both frames are related via a Lorentz boost, which leaves the
$x$ component unchanged, leading to
\beq
\label{11e}
p \sin \theta = p_0 \sin \theta_0\ .
\eeq

\noi
On the other hand $\delta \varphi \equiv \delta \varphi_0$,
 since the Lorentz boost does not affect the projection on the $zx$ plane.
 Using eqs. (\ref{4e}) and (\ref{5e}), we also find~:
\beq
\label{12e}
{\delta \theta \over \sin \theta} = {\delta \theta_0 \over \sin \theta_0} \ .
\eeq

\newpage
\subsection{Comparing frames}
\hspace*{\parindent}

The corresponding transverse momentum broadening is easily obtained
 in both frames~: we sum the squares of the momentum increase in the $z$
 axis direction and in the direction orthogonal to the parton momentum~:
\beq
\label{13e}
\delta p_{0\bot}^2 = \delta p_{0z}^2 + \delta p{'}^2_0 \equiv p_0^2
 \left [ \sin^2 \theta_0 \ \delta \varphi^2 + \delta \theta_0^2 \right ] \ .
\eeq

\noi Similarly
\beq
\label{14e}
\delta p_{\bot}^2 = p^2 \left [ \sin^2 \theta \ \delta \varphi^2 + 
\delta \theta^2 \right ] \ .
\eeq

\noi Using eqs.~(\ref{11e}) and (\ref{12e}), we find that
 $\delta p_{\bot}^2 = \delta p_{0\bot}^2$ and may immediately infer
 the relation between $\widehat{q}_0$ and $\widehat{q}$~: indeed we get
\bea
\label{15e}
\delta p_{\bot}^2 &=& \widehat{q} \ \delta \tau \qquad \hbox{and} \\
\delta p_{0\bot}^2 &=& \widehat{q}_0 \ \delta \tau_0 =
 \widehat{q}_0 \ \delta \tau \left [ {\rm ch}\
 \zeta \left ( 1 + \cos \theta \ {\rm th}\ \zeta\right ) \right ] \ ,
\label{16e}
\eea

\noi leading to (locally)~:
\beq
\label{17e}
\widehat{q} \equiv \widehat{q}_0\ {\rm ch}\ \zeta \left ( 1 + \cos
 \theta ~{\rm th} \zeta \right )  \ .
\eeq

\noi Recently
this transformation property has  been independently derived
in \cite{Urs}, and already used in \cite{Renk:2006sx}.

\noi $\widehat{q}_0$ may be parametrized as discussed in the Introduction.

\noi Writing the transverse flow velocity as $u^{\mu} = ({\rm ch}\ \zeta ,
 0,  - {\rm sh}\ \zeta , 0)$ and with the parton momentum written as
 $\displaystyle{{p^{\mu} \over p_0}} = 
(1, \sin \theta,  \cos \theta, 0 )$, we get
\beq
\label{18e}
\widehat{q} = \widehat{q}_0 \ {u^{\mu} \cdot p^{\mu} \over p_0}\ .
\eeq

\noi A way to understand this relation is to write the transport
 coefficient as the rate of interaction -- the number of interactions
 per unit time -- multiplied by the typical invariant scale
 $\mu^2$ characterizing the momentum transfer to the parton
 in a parton-medium collision~: for a medium at rest
 $\widehat{q}_0 = R_0 \mu^2$ and for a moving medium
 $\widehat{q} = R \mu^2$. The rate of interaction varies indeed like the 
inverse of the time spent in the medium~:
\beq
\label{18ea}
{R \over R_0} = {\rm ch} \ \zeta \left ( 1 + \cos \theta \ {\rm th}
 \ \zeta \right )  \ ,
\eeq

\noi so that eq.~(\ref{17e}) is recovered.

\subsection{Parton transverse broadening}
\hspace*{\parindent}

\noi Using $\widehat{q}_0 (\tau )$
we may now calculate the transverse broadening 
\cite{Baier:1996sk} obtained
 when taking into account the transverse flow for a parton 
travelling in a thermalized medium at temperature $T$.
 This last condition allowing
 us to determine the $\tau$ dependence of $T$ and the transverse velocity 
$v$ by using hydrodynamical model codes. \par

\begin{figure}[t]
\centerline{\epsfig{bbllx=1,bblly=1,bburx=460,bbury=330,
file=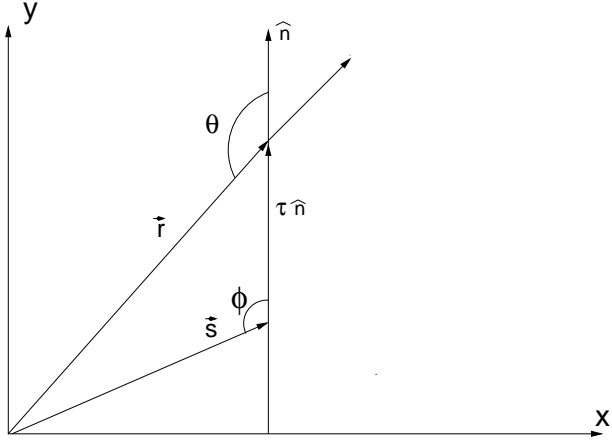,width=80mm}}
\caption{\label{fig:N0}\vspace{0.3cm} View of the kinematics in the transverse plane. 
}
\end{figure}

In Fig.~\ref{fig:N0} we have schematized the situation
 in the transverse plane \cite{Eskola:2004cr}~:
 the vector $\vec{s} (s, \phi )$ defines the point in the plane
 where the hard parton enters the medium. The parton then follows a
 path along the unit vector $\widehat{n}$ and along its way,
 encounters the radial flow along the direction $\vec{r}$ which
 depends on time $\tau$.\par

\vspace{0.3cm}

\noi
We write $\vec{r} = \vec{s} + \tau \widehat{n}$ and defining 
 $\theta$ (according to the definition given in sect.~2) 
and $\phi$ as in Fig.~\ref{fig:N0},  we find that
\beq
\label{19e}
\cos \theta =  { s \cos \phi - \tau \over \sqrt{s^2 + \tau^2 - 2 s
 \tau \cos \phi}} \ .
\eeq

\noi Writing ${\rm ch}\  \zeta$ as $\displaystyle{{1
 \over \sqrt{1 - v^2}}} \equiv \gamma (v)$ and 
${\rm th}\  \zeta = v$, 
where $v$ depends on $r$ and $\tau$. We may now get the following expression
 for the transverse broadening of the parton for a cylindrical medium at
 central rapidity~:
\bea
\label{20e}
(\Delta p_{\bot}^2)_{Bj + flow} &=&
 {1 \over \pi R_A^2} \int_0^{R_A} sds \int_0^{2\pi}
 d\phi \int_{\tau_0}^{\tau_{max}} d\tau\ \widehat{q}_0 \left ( T (r,
 \tau )\right ) \gamma (v) \left ( 1 + v \cos \theta \right ) 
\eea

\noi where $\tau_{max}$ is defined as the maximum in-medium path
 length compatible with geometry and taking into account the life-time
 of the thermalized medium $\tau_{QGP}$~:
\beq
\label{21ea}
\tau_{max} = \ {\rm min} \left \{ \tau_{max}^{geom}, \tau_{QGP}\right \} \ ,
\eeq

\noi with
\beq
\label{21e}
\tau_{max}^{geom} =  s \cos \phi + \sqrt{s^2 \cos^2 \phi + R_A^2 - s^2} \ ,
\eeq

\begin{figure}[t]
\vspace{2.0cm}
\centerline{\epsfig{bbllx=1,bblly=1,bburx=600,bbury=460,
file=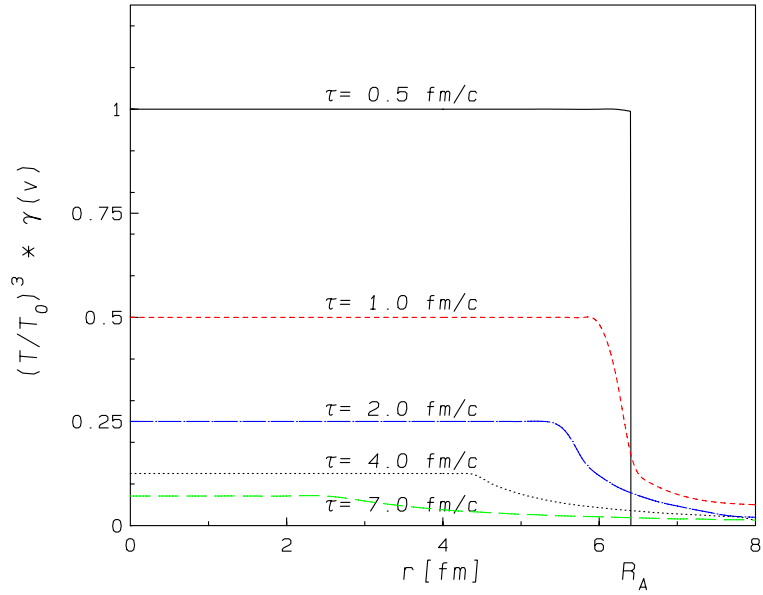,width=100mm}}
\caption{\label{fig:N2}  $(T/T_0)^3\cdot \gamma (v)$ as a function of $r$
for different values of $\tau$ calculated from ideal hydrodynamics 
\cite{Baym}.
}
\end{figure}

\noi and $\tau_0$ is the initial time.
In the integral (\ref{20e}) the boundaries are simplified by neglecting the
flow beyond the nucleus radius $R_A$ (c.f. Fig.~\ref{fig:N2}).
 The life-time $\tau_{QGP}$ is
 estimated in the framework of Bjorken expansion \cite{Bjorken}, after having fixed the
 initial temperature and taking the final temperature as $T \simeq 200$~MeV.
 (In the following we take $\tau_{QGP} \simeq 4.5 ~$fm).
\par

We  actually calculate the ratio 
\beq
\label{200e}
R_{flow} =
 (\Delta p_{\bot}^2)_{Bj + flow} /(\Delta p_{\bot}^2)_{Bj}
\eeq 
in order to quantify the importance of the radial flow in addition to
 Bjorken ($Bj$) longitudinal expansion \cite{Bjorken},
for which
\beq
\label{Bje}
\hat{q_0} (T) = \hat{q_0}(T_0) {(\frac{T}{T_0})}^3 = 
\hat{q_0}(\tau_0) (\frac{\tau_0}{\tau})^{3c_s^2} \ ,
\eeq
where $c_s$ denotes the sound velocity (taken to be $c_s =
\frac{1}{\sqrt{3}}$).

\begin{figure}[t]
\vspace{-0.5cm}
\centerline{\epsfig{bbllx=1,bblly=80,bburx=700,bbury=760,
file=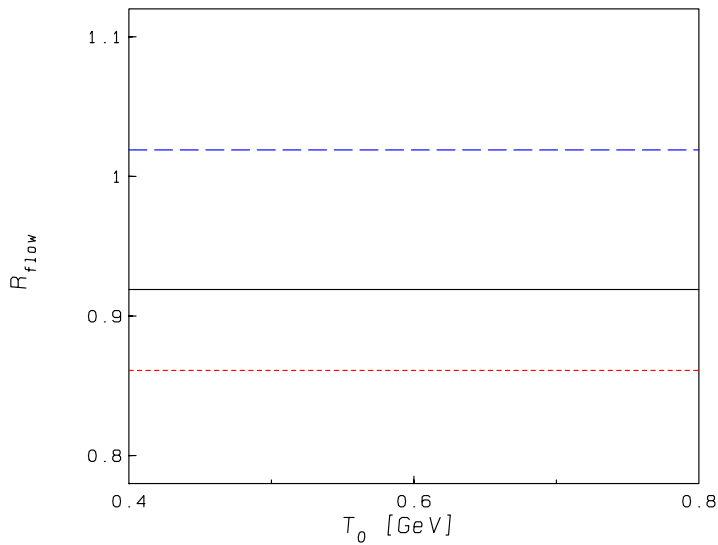,width=95mm}}
\caption{\label{fig:N1} $R_{flow}$ as a function of initial temperature $T_0$
from eqs.~(\ref{20e}) and (\ref{200e}) (solid curve). Dashed and   
dotted curves as described in the text.}
\end{figure}

\noi  The calculation is
 done in the framework of ideal hydrodynamics,
assuming initial conditions relevant for heavy-ion collisions, e.g.
a vanishing flow velocity at initial time $\tau_0$ \cite{Baym}.
 The result is shown as the solid curve in
 Fig.~\ref{fig:N1}
 for a cylindrical medium of radius $R_A = 6.4$~fm ($Au$ nucleus), 
fixing the initial time $\tau_0 = 0.5$~fm, as a function of
 the initial temperature $T_0$. \par
The curves are actually calculated using the approximate analytic expressions
derived in \cite{Baym} (see also \cite{Baier:2006gy}).
In the ratio $R_{flow}$ the dependence on the initial temperature $T_0$ cancels. 
Ignoring the limit given by $\tau_{QGP}$ results in the dotted curve.
We note that $R_{flow}$ is smaller than 1. This can be understood in the
 following way: the flow has a non
negligible effect  for large enough values of $r$, where $v$  
differs significantly from 0.
 But this is only realized (see Fig.~\ref{fig:N0})  when the jet is
moving   with the flow, since then  $\cos \theta  \simeq -1$, and
therefore $\gamma (v) (1+ v \cos \theta) < 1$.
This effect is even larger for the dotted curve.

 We note that the angular dependence 
of eq. (\ref{17e}), which is responsible
for the reduction of broadening, and also quenching,
is not present in the ansatz for $\widehat{q}$ given
in \cite{Armesto:2004vz} and used in \cite{Renk:2005rq,Renk:2005ta}.

As an exercise we calculated $R_{flow}$ with the following set-up:
we assume that $\widehat{q_0}(T,\tau)$ follows Bjorken's $\tau$ dependence
(\ref{Bje}), keeping the hydro velocity $v(r,\tau)$ in $\gamma (v)$, but without 
the $r$-dependence in the temperature, and dropping the $\cos \theta$ term.
The ratio  $R_{flow}$
becomes larger than one, which can be seen from the dashed curve
 in Fig.~\ref{fig:N1},
because in this case the factor $\gamma (v)$ wins.

Two conclusions emerge~:

a) the effect of the transverse flow is small and of the order $\le 10\ \%$.

b) the ratio $ R_{flow}$ is smaller than 1 which goes against the possibility 
that the transverse flow would allow us to solve the
 perturbative/non-perturbative dilemma discussed in the introduction. \par

The smallness of the effect is due to the fact that the
 transverse velocity takes a long time to reach an appreciable value,
 starting from $v = 0$ and cannot provide the possibility of
 compensating the cooling effect. This can be visualized in Fig.~\ref{fig:N2}
 where we have plotted the quantity $(T/T_0)^3\cdot \gamma (v)$ --
 which is essentially the integrand in eq. (\ref{20e}) --
 for different values of $\tau$, as a function of $r$, using ideal
 hydrodynamics \cite{Baym}.

\begin{figure}[t]
\vspace{2.0cm}
\centerline{\epsfig{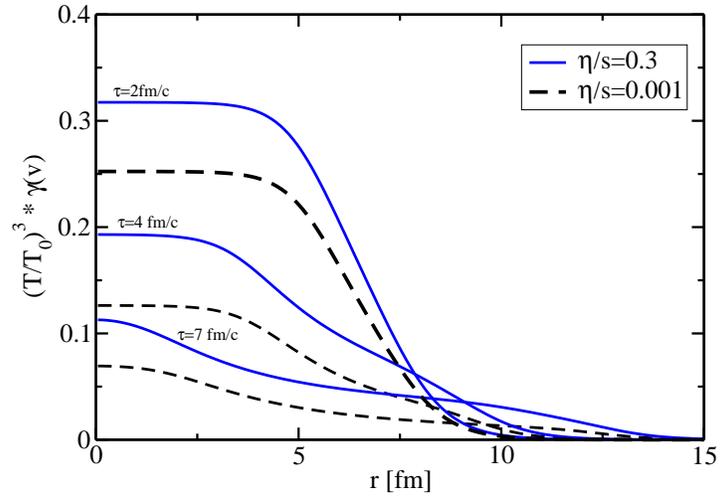}}
\caption{\label{fig:N3} Effects due to non-vanishing shear viscosity $\eta$
\cite{Paul}.}
\end{figure}

It is interesting to investigate the 
effects due to a non-vanishing shear viscosity  compared to the ideal
hydrodynamical description for 
the quantity $(T/T_0)^3\cdot \gamma (v)$, which  
is plotted in Fig.~\ref{fig:N3}. The details of this
calculation may be found in \cite{Baier:2006gy}.

No significant effects due to the transverse flow in the presence of
shear viscosity are found, as maybe  expected, when comparing the solid
 and dashed curves in
Fig.~\ref{fig:N3}, although cooling is slowed down at least
for times $\tau$ presented in this figure: $R_{flow}$ stays of $O(1)$.

\section{Energy loss}

The medium-induced radiative jet energy loss is  determined by the
transport coefficient $\hat{q}$.
While the energy loss formalism has been previously developed only for
a medium without flow it is apparent from the discussion at the end of
section 3.B that if one uses $\widehat{q}$, as given in eq.~(\ref{17e}),
to determine the interaction of QCD partons with the medium then there
is no modification of the fundamental energy loss formulas.
Thus eqs.~(25) and (26) of \cite{Baier:1998yf} remain valid in the
 presence of flow.
 The basic quantity
is the radiation spectrum of gluons emitted from the high energy parton.
In the soft gluon energy limit, $\omega$
much smaller than the parton energy, the dominant mechanism consists in
multiple scatterings of the radiated gluon off the "centers" describing
the medium \cite{Baier:1996kr}. For static centers, as well as for a dense (thermal)
 medium which undergoes
longitudinal Bjorken-type expansion \cite{Bjorken}
 the gluon spectrum
obeys the following scaling law in terms of a characteristic
gluon energy $\omega_c$ \cite{Salgado:2002cd,Kovner:2003zj},

\bea
\label{40e}
\omega \frac{dI}{d\omega} = \tilde{I}(\omega/\omega_c)~.  
\eea

\noi
For static centers \cite{Baier:1996kr} 

\bea
\label{41e}
\omega_c = \frac{1}{2} \hat{q} L^2~,
\eea
 where $L$ denotes the path length of the energetic jet in the medium. 

\noi
In the expanding case \cite{Baier:1998yf}, eq.~(\ref{41e}) is generalized to
\cite{Salgado:2002cd,Wang:2002ri}

\bea
\label{42e}
\omega_c = \int_0^L ~d\tau \tau \hat{q}(\tau) ~,
\eea
with $\hat{q}(\tau) \approx 1/\tau$, when following Bjorken \cite{Bjorken}.

However, when transverse flow is present in the medium this nice scaling
property does not hold, since $\hat{q}$ appears at different times,
corresponding to the interference product of the emission amplitude
 and the complex
conjugate one (see eg. eq.~(25) in \cite{Baier:1998yf}).
As for the case of $p_\perp$-broadening,
we do not expect large effects due to transverse flow. 
Therefore, for simplicity, we calculate the ratio

\bea
\label{43e}
R_{\omega_c} = \frac{(\omega_c)_{Bj + flow}}{(\omega_c)_{Bj}}~.
\eea
In comparison with the calculation of $R_{flow}$, we only estimate
 the first moment
with respect to the path length  $\tau$ of the energetic jet, i.e.

\bea
\label{44e}
(\omega_c)_{Bj + flow} =
 {1 \over \pi R_A^2} \int_0^{R_A} sds \int_0^{2\pi} 
d\phi \int_0^{\tau_{max}} d\tau\ \tau\ \widehat{q}_0 
\left ( T (r, \tau )\right ) \gamma (v)
 \left ( 1 + v \cos \theta \right ) \ ,
\eea
and accordingly for $(\omega_c)_{Bj}$.

\noi As we expect from Fig.~\ref{fig:N1} the ratio $R_{\omega_c}$
does not differ significantly from  $1$, when calculated within
(ideal) hydrodynamics and realistic initial
conditions for the flow field
$v(r, \tau_0)$: a typical value is $R_{\omega_c} \simeq
0.85$ for $\tau_0 = 0.5~fm, T_0 = 400 ~MeV$.
From this estimate, we expect small effects of radial flow
on the quenching of large transverse momentum hadrons produced in
nucleus-nucleus collisions. 
This conclusion on $R_{\omega_c} < 1$
 is confirmed in the recent analysis in \cite{Renk:2006sx}
using eq.~(\ref{17e}).

\noi
Our study seems to  rule out the possibility of reducing the 
large value of $\widehat{q}$ from model comparisons with data
\cite{Eskola:2004cr} by
including radial flow with initial conditions,
which we assume to be  realistic and causal in heavy-ion collisions.

\vspace{0.5cm}

{\bf Acknowledgements}

We would like to thank P.~Romatschke and 
 U.~A.~Wiedemann for stimulating discussions
and useful comments. 
We thank K.~Rajagopal for questions and T.~Renk for 
comments, which led us to revise our
original paper.
This work was partially supported by the US
Department of Energy.


\begin{thebibliography}{99}
\bibitem{Paul}
We kindly thank Paul Romatschke for this figure.
\end{thebibliography}
\end{document}